%% file: main.tex
\PassOptionsToPackage{hyphens}{url}

\documentclass[sigconf]{acmart}

\setcopyright{none}
\acmDOI{}
\acmISBN{}

\usepackage{microtype}
\hypersetup{breaklinks=true}
\usepackage{tabularx}
\usepackage{booktabs}
\PassOptionsToPackage{hyphens}{url}
\usepackage{xurl}
\usepackage{tcolorbox}
\usepackage{enumitem}
\usepackage{tikz}
\usetikzlibrary{arrows.meta,positioning,shapes.geometric}
\usepackage{pgfplots}
\usepackage{pgfplotstable}
\pgfplotsset{compat=1.18}
\usepackage{xcolor}
\usepackage{graphicx}
\graphicspath{{sections/figures/}} 
\usepackage{tabularx}
\usepackage{makecell} %

\copyrightyear{2026}
\acmYear{2026}
\setcopyright{cc}
\setcctype{by}
\acmConference[MSR '26]{23rd International Conference on Mining Software Repositories}{April 13--14, 2026}{Rio de Janeiro, Brazil}
\acmBooktitle{23rd International Conference on Mining Software Repositories (MSR '26), April 13--14, 2026, Rio de Janeiro, Brazil}
\acmDOI{10.1145/3793302.3793561}
\acmISBN{979-8-4007-2474-9/2026/04}

\begin{document}
\title[Collaboration Signals Shaping the Integration of Agent-Authored Pull Requests]{When AI Teammates Meet Code Review: Collaboration Signals Shaping the Integration of Agent-Authored Pull Requests
}

\author{Costain Nachuma}

\affiliation{%
  \institution{Idaho State University\country {USA}}
  }
  \email{costainnachuma@isu.edu}

\author{Minhaz F. Zibran}
\affiliation{%
  \institution{Idaho State University\country {USA}}
  }
  \email{zibran@isu.edu}


\begin{abstract}
\input{sections/abstract}
\end{abstract}





\maketitle

\input{sections/introduction}

\section{Dataset and Operational Definitions}\label{sec:data}

\input{sections/data}

\input{sections/RQ1}

\input{sections/RQ2}

\input{sections/discussion}

\input{sections/related}

\input{sections/threats}

\input{sections/conclusion}

\bibliographystyle{ACM-Reference-Format}
\bibliography{references}

\end{document}

%% file: sections/abstract.tex
Autonomous coding agents increasingly contribute to software development by submitting pull requests on GitHub; yet, little is known about how these contributions integrate into human-driven review workflows. We present a large empirical study of agent-authored pull requests using the public AIDev dataset, examining integration outcomes, resolution speed, and review-time collaboration signals. Using logistic regression with repository-clustered standard errors, we find that reviewer engagement has the strongest correlation with successful integration, whereas larger change sizes and coordination-disrupting actions, such as force pushes, are associated with a lower likelihood of merging. In contrast, iteration intensity alone provides limited explanatory power once collaboration signals are considered. A qualitative analysis further shows that successful integration occurs when agents engage in actionable review loops that converge toward reviewer expectations. Overall, our results highlight that the effective integration of agent-authored pull requests depends not only on code quality but also on alignment with established review and coordination practices.

%% file: sections/introduction.tex
\section{Introduction}

Autonomous coding agents are increasingly participating in everyday software development ~\cite{wessel2018power}. Beyond acting as code generation tools, these systems now submit pull requests that fix bugs, add features, and extend test suites on GitHub. As a result, human developers are routinely asked to review, carefully interpret, and integrate changes authored by non-human contributors\cite{cognition2024devin,li2025aidev}.

Recent large-scale studies based on the AIDev dataset document the rapid growth of agent-authored pull requests and their aggregate merge rates across projects \cite{aidev_zenodo_v3,li2025aidev}.  However, we note that adoption alone does not imply effective collaboration. In pull-based development, integration outcomes reflect not only the technical correctness of a change but also how contributors engage with review workflows, iteration cycles, and coordination norms \cite{gousios2014pullbased,rigby2014peer,tsay2014discussion}. Whether autonomous agents can function as effective collaborators within these socio-technical processes, rather than merely producing correct patches, remains unclear.

Two gaps motivate our study. First, while agent-authored pull requests are becoming increasingly common, we lack a clear understanding of how reliably they integrate and how long it takes to reach a decision. Second, beyond agent identity, there is limited empirical evidence on which review-time collaboration signals, such as iteration behavior, reviewer engagement, testing activity, and coordination stability, are associated with successful integration. Despite extensive prior work on pull-based development \cite{gousios2014pullbased, rigby2014peer, tsay2014discussion}, there is limited empirical evidence on how autonomous agents engage with these review-time collaboration processes.

Addressing these gaps is essential as software development shifts toward mixed human–AI teams. For agents to serve as effective collaborators rather than isolated patch generators, they must align with the implicit social and procedural expectations of code review. In this paper, we present a large-scale empirical analysis of agent-authored pull requests, focusing on how review-time collaboration dynamics shape integration outcomes. Rather than treating these pull requests solely as code artifacts, we analyze their integration as a socio-technical process mediated by human review and coordination. We examine the following two research questions:

\vspace{0.1cm}
\noindent \textbf{RQ1:} \textit{Among agent-authored pull requests, what proportions are merged, closed without merging, or remain open, and how long do resolved pull requests take to reach a decision?} \\
\emph{Measures:} share(merged), share(closed-unmerged), share(open), and time-to-decision (mean and median).

\vspace{0.1cm}
\noindent \textbf{RQ2:} \textit{Which review-time collaboration signals are associated with the integration of agent-authored pull requests?} \\
\emph{Focus:} iteration intensity, testing behavior, reviewer engagement, and coordination stability (from GitHub review-time traces). \\
\emph{Signals:} number of commits (bucketed), $\Delta$LOC and number of files changed, a tests-added indicator, a force-push indicator, review presence, and time to first review.


%% file: sections/data.tex
\begingroup
\sloppy

\noindent\textbf{Dataset snapshot.}
We use the \textbf{AIDev} dataset of agent-authored pull requests (PRs) from Zenodo
(version~3; accessed Nov~2025), citing both the dataset record and its accompanying
preprint \cite{aidev_zenodo_v3,li2025aidev}.

\noindent\textbf{Acquisition and binding.}
All Parquet tables were obtained from Zenodo version~3 and bound as DuckDB views
using an in-memory DuckDB engine. The tables used include
\textit{pull\_request}, \textit{pr\_task\_type}, \textit{pr\_commits},
\textit{pr\_commit\_details}, \textit{pr\_reviews}, \textit{pr\_timeline},
\textit{repository}, and \textit{user}.

\noindent\textbf{Dataset scope (curated popular).}
Our analysis focuses on the curated subset of popular repositories with at least
100 GitHub stars, following the dataset authors’ recommended filtering.
This subset contains 33{,}596 agent-authored PRs submitted by 1{,}797 distinct
developers across 2{,}807 repositories, spanning multiple coding agents.

\noindent\textbf{Agent-authored PRs.}
A PR is considered agent-authored if it appears in \textit{pr\_task\_type} with a
non-null \texttt{agent} label. Unless otherwise noted, all reported statistics and
analyses are computed over this set. The dataset does not distinguish between autonomous and human-invoked agent submissions; we therefore treat all agent-authored pull requests uniformly.

\noindent\textbf{Outcomes and timestamps.}
PR outcomes are derived directly from GitHub timestamps to avoid ambiguity:
\[
\begin{aligned}
\texttt{merged} &:= (\texttt{merged\_at} \neq \varnothing), \\
\texttt{closed\_without\_merge} &:= (\texttt{closed\_at} \neq \varnothing \ \wedge\ \texttt{merged\_at} = \varnothing), \\
\texttt{open} &:= (\texttt{closed\_at} = \varnothing \ \wedge\ \texttt{merged\_at} = \varnothing).
\end{aligned}
\]
For resolved PRs, the decision time is taken as the merge time when present and
otherwise the closure time. All timestamps are normalized to UTC.

\noindent\textbf{Time to decision.}
We compute time to decision as the elapsed time between \texttt{created\_at} and the
decision timestamp for resolved PRs. Due to right-skewed lifetimes, we report both
mean and median values.

\noindent\textbf{Measures.}
We report the share of agent-authored PRs that are merged, closed without merge,
or remain open. We use Wilson score confidence intervals (95\%), which provide
more reliable coverage than normal approximations for binomial proportions,
particularly near boundary values \cite{agresti-coull-1998,newcombe-1998}.

\noindent\textbf{Replication package.}
We release a complete replication package containing all scripts required to reproduce our results~\cite{repkg_aidev_msr2026_anon}. 
RQ1 is reproduced by running \verb|python run_rq1.py --config config.yaml|. 
RQ2 is reproduced by first constructing features and then fitting regression models using the provided scripts.

\endgroup

%% file: sections/RQ1.tex
\section{Integration and Resolution (RQ1)}
\label{sec:rq1}

\subsection{Motivation}
Pull requests (PRs) are the dominant mechanism for integrating changes in modern, pull-based development, and prior work has shown that PR outcomes (merge vs. non-merge) reflect both technical and social processes during review \cite{gousios2014pullbased,rigby2014peer,tsay2014discussion}. As autonomous coding agents increasingly submit PRs at scale \cite{aidev_zenodo_v3,li2025aidev}, it is important to establish a baseline: (i) whether agent-authored PRs integrate at rates comparable across agents, and (ii) how quickly they reach a decision. This baseline anchors subsequent analyses.

\subsection{Approach}
We study agent-authored PRs from the AIDev dataset \cite{aidev_zenodo_v3,li2025aidev}. For each PR, we derive its outcome directly from GitHub timestamps to avoid ambiguity in API state fields: a PR is \emph{merged} if \texttt{merged\_at} is present; \emph{closed-unmerged} if \texttt{closed\_at} is present and \texttt{merged\_at} is absent; otherwise it is \emph{open}. We report outcome shares overall and per agent, along with 95\% confidence intervals for proportions using standard interval estimation methods \cite{newcombe-1998,agresti-coull-1998}.

To characterize resolution dynamics, we compute \emph{time-to-decision} for resolved PRs (merged or closed-unmerged) as the elapsed time from \texttt{created\_at} to \texttt{merged\_at} (if merged) or \texttt{closed\_at} (if closed-unmerged). We summarize decision latency using both the mean and median to account for skew common in PR lifetimes.

\vspace{-0.2cm}
\paragraph{Exploratory note on reverts.}
Reverts can be an imperfect proxy for post-merge problems, as repository-level signals may arise from refactoring, redesign, or other maintenance activities rather than defects \cite{herzig2013szz}. Accordingly, we do not consider reverts as a primary outcome and focus our analysis on review-time integration and collaboration dynamics.


\input{sections/figures/distribution}

\subsection{Results}
Across \textbf{33,596} agent-authored PRs spanning \textbf{5} coding agents, \textbf{71.5\%} were merged, \textbf{21.6\%} were closed without merge, and \textbf{6.9\%} remained open at the time of data collection (merged share 95\% CI: \textbf{[0.710, 0.720]}). These aggregates indicate that most agent-authored PRs reach integration, but a substantial fraction still fail to merge or remain unresolved.

\vspace{-0.2cm}
\subsubsection{Outcome rates vary sharply by agent.}
Merge rates differ markedly across agents, suggesting heterogeneity in both PR quality and fit to reviewer expectations. For example, OpenAI\_Codex shows the highest merge share (\textbf{82.6\%}), while Copilot merges substantially less often (\textbf{43.0\%}); Devin lies in between (\textbf{53.8\%}). Importantly, the \emph{open} share also differs across agents (e.g., Copilot exhibits a notably higher open fraction than others), indicating different resolution dynamics and/or triage practices.

\vspace{-0.2cm}
\subsubsection{Decision latency differs even more than merge rates.}
Resolution speed varies substantially by agent. For resolved PRs, OpenAI\_Codex reaches a decision extremely quickly (median \textbf{$<$1 hour}; mean $\approx$ \textbf{19.4 hours}), while Copilot and Devin are slower (median \textbf{13 hours} and \textbf{9 hours}, respectively; with mean times exceeding \textbf{80--100 hours}). This pattern suggests differences in review effort and alignment with repository norms.

\subsection{Summary}
RQ1 establishes a baseline for agentic PR integration and resolution dynamics. Overall, a majority of agent-authored PRs merge, but integration success and decision speed vary strongly across agents. These differences motivate our subsequent analyses of review-time collaboration signals: beyond whether an agent can produce working code, the interaction and coordination costs borne by human reviewers may be a key determinant of whether and how quickly agent-authored PRs integrate \cite{gousios2014pullbased,tsay2014discussion,rigby2014peer}.
\vspace{-5px}
\begin{tcolorbox}[boxrule=0.5pt, boxsep=-2pt, left=5pt, right=5pt]
\textbf{Answer to RQ1:}
Most agent authored pull requests are successfully merged, but both integration rates and resolution speed vary substantially across agents. These differences indicate that agentic PR integration depends not only on code correctness, but also on how agents interact with human review workflows and manage coordination costs during review.
\end{tcolorbox}

%% file: sections/figures/distribution.tex
\begin{figure}[t]
\centering
\begin{tikzpicture}
\begin{axis}[
    ybar stacked,
    width=0.80\linewidth,
    height=4.2cm,
    bar width=14pt,
    ymin=0, ymax=100,
    ylabel={Share of PR outcomes (\%)},
    symbolic x coords={
        OpenAI\_Codex,
        Copilot,
        Devin,
        Cursor,
        Claude\_Code
    },
    xtick=data,
    xticklabel style={rotate=45, anchor=east},
    enlarge x limits=0.15,
    ymajorgrids=true,
    grid style=dashed,
    legend style={
        at={(0.5,1.05)},
        anchor=south,
        legend columns=-1,
        fill=none,
        draw=none,
        font=\footnotesize      
    },
    tick label style={font=\small},
    label style={font=\small}
]

\addplot+[fill=blue!70] coordinates {
    (OpenAI\_Codex,83)
    (Copilot,43)
    (Devin,54)
    (Cursor,65)
    (Claude\_Code,59)
};

\addplot+[fill=orange!80] coordinates {
    (OpenAI\_Codex,14)
    (Copilot,35)
    (Devin,43)
    (Cursor,22)
    (Claude\_Code,24)
};

\addplot+[fill=green!70] coordinates {
    (OpenAI\_Codex,4)
    (Copilot,22)
    (Devin,3)
    (Cursor,13)
    (Claude\_Code,17)
};

\legend{Merged, Closed (unmerged), Open}

\end{axis}
\end{tikzpicture}
\vspace{-0.4cm}
\caption{Outcomes of agent-authored pull requests}
\vspace{-0.3cm}
\end{figure}
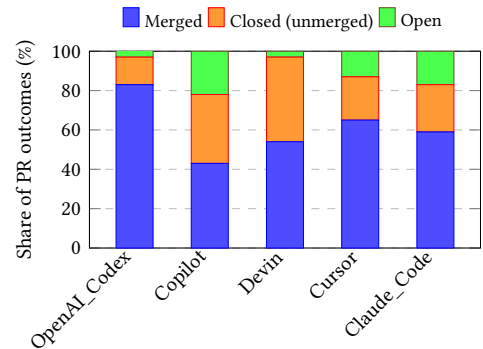

%% file: sections/RQ2.tex
\section{Collaboration Signals (RQ2)}
\label{sec:rq2}

\subsection{Motivation}
RQ1 revealed substantial variation in integration outcomes and decision latency across coding agents. Beyond agent identity, this raises the question of which \emph{review-time collaboration signals} are associated with the successful integration of agent-authored pull requests (PRs). Prior work shows that reviewers rely on both technical and social cues, such as change size, reviewer participation, and coordination stability when evaluating pull requests \cite{gousios2014pullbased,tsay2014discussion,rigby2014peer}. We examine whether similar signals shape the acceptance of agent-authored PRs.

\subsection{Approach}
We model pull request integration as a binary outcome (merged vs.\ closed without merge). We focus on \emph{review-time collaboration signals} observable in GitHub traces prior to the final decision, capturing how agent-authored pull requests align with review workflows through iteration and change magnitude, coordination stability, and reviewer engagement. We estimate interpretable associations between these signals and merge decisions using logistic regression. Predictors capture three dimensions: (1) iteration and change magnitude, (2) coordination stability and process signals, and (3) reviewer engagement. Agent indicators are included to control for systematic differences across coding agents.

Change magnitude is measured using log-transformed code churn ($\log(1+\Delta\text{LOC})$) and the number of files modified. Coordination stability is captured via a binary indicator for force pushes during review. Testing behavior is represented by a binary indicator flagging whether test files were modified. Reviewer engagement is measured using indicators for whether a pull request received at least one review and the log-transformed time to first review. Iteration intensity is operationalized either through commit-count buckets (Model~A) or log-transformed commit counts (Model~B), following prior work on code review and pull request integration~\cite{mcintosh2014impact}.

To address skew, size-related variables are log-transformed using $\log(1+x)$. We estimate logistic regression models with repository-clustered standard errors and report odds ratios. Results are interpreted as associations rather than causal effects. Claude Code is used as the reference category for agent indicators.

\subsection{Results}
\input{sections/figures/forestPlot}

Figure~\ref{fig:forestplot} reports odds ratios from logistic regression models. Model~A uses commit-count buckets, while Model~B uses log-transformed commit counts to operationalize iteration intensity. Consistent effects across models indicate robustness. The x-axis shows odds ratios on a logarithmic scale.


\vspace{-0.2cm}
\subsubsection{Reviewer engagement dominates integration outcomes.}
Across both models, receiving at least one review is the strongest correlate of merging. Agent-authored PRs that receive reviewer attention exhibit substantially higher odds of integration, suggesting that reviewers selectively invest effort in changes perceived as viable.

\vspace{-0.2cm}
\subsubsection{Coordination-disrupting behaviors are penalized.}
Force pushes during review are consistently associated with lower merge likelihood; while sometimes required for repository hygiene (e.g., squash or rebase policies), history rewriting during active review can disrupt shared understanding and increase coordination costs for reviewers.

\vspace{-0.2cm}
\subsubsection{Change size increases review burden.}
Larger changes are less likely to be merged even after controlling for agent identity and reviewer engagement. This aligns with prior PR research and suggests that reviewers apply similar risk heuristics to agent- and human-authored contributions.

\vspace{-0.2cm}
\subsubsection{Iteration and testing show limited independent effects.}
Neither iteration intensity nor test additions are significantly associated with merge outcomes once reviewer engagement and coordination stability are accounted for. Producing more revisions or adding tests alone does not increase integration likelihood without reducing reviewer uncertainty.

\vspace{-0.2cm}
\subsubsection{Time-to-review reflects workflow dynamics.}
Conditional on receiving a review, longer time to first review is positively associated with merging, likely reflecting repository-level prioritization rather than a direct benefit of delayed review.

\vspace{-0.2cm}
 \subsubsection{Synthesis.}
Overall, reviewers reward agent behaviors that reduce coordination cost and penalize those that disrupt shared understanding. Successful integration depends less on activity volume and more on alignment with the social and procedural norms of code review. Comparing these patterns with fully human-authored pull requests is an important direction for future work.

\vspace{-2px}

\subsection{Qualitative Follow-up}
While the regression analysis identifies which review-time collaboration signals are associated with integration outcomes, it does not explain how these signals manifest during review. We therefore analyzed a random and purposive sample of 60 agent-authored pull requests spanning merged and non-merged outcomes. Using a predefined codebook derived from prior pull request review literature and pilot inspection of agent-authored review threads, each pull request was assigned a single primary code capturing the dominant review-time mechanism influencing its outcome. Disagreements were resolved through discussion. Table~\ref{tab:rq2_qualitative_summary} summarizes the results.

\vspace{-0.2cm}
\subsubsection{Actionable review convergence enables success.}
The dominant success pattern is an \emph{actionable review loop}, accounting for 32 PRs and 30 of all merged cases. In these cases, reviewers provided concrete feedback and agents responded with targeted revisions that converged toward acceptance, explaining the strong positive association between reviewer engagement and merging.

\vspace{-0.2cm}
\subsubsection{Design conflict and coordination breakdown explain rejection.}
All non-merged PRs are associated with failure mechanisms. Design or architectural disagreements reflect misalignment with project principles, while coordination breakdowns (e.g., force pushes) disrupt review context. These patterns directly correspond to negative associations observed in the regression models.

\vspace{-0.2cm}
\subsubsection{Incomplete or policy-blocked changes rarely recover.}
PRs coded as incomplete solutions or blocked by process constraints (e.g., policy or repository state) consistently failed to merge, clarifying why iteration alone does not improve outcomes once engagement and coordination stability are considered.

\vspace{-0.2cm}
\subsubsection{Iteration without convergence is insufficient.}
Consistent with quantitative results, additional commits or test additions do not lead to integration unless they directly address reviewer concerns and reduce uncertainty.

\vspace{-0.2cm}
\subsubsection{Synthesis.}
Taken together, the qualitative findings show that reviewer engagement is effective only when agents participate in stable, goal-aligned review interactions. Integration success depends on convergence through feedback rather than activity volume.
\vspace{-2px}
\begin{tcolorbox}[boxrule=0.5pt, boxsep=-2pt, left=5pt, right=5pt]
\textbf{Answer to RQ2:}
Integration of agent-authored pull requests is driven primarily by review-time collaboration signals rather than iteration volume. Reviewer engagement increases merge likelihood when it produces actionable feedback and convergence, while large changes and coordination-disrupting behaviors reduce success. Iteration alone has limited impact without alignment with reviewer expectations and project norms.
\end{tcolorbox}

\input{sections/figures/rq2_qualitative}


%% file: sections/figures/forestPlot.tex
\setlength{\textfloatsep}{8pt plus 2pt minus 2pt}
\begin{figure}[t]

\centering
\begin{tikzpicture}

\begin{axis}[
    width=\linewidth,
    height=5.0cm,                 
    xmode=log,
    xmin=0.1, xmax=10,
    xlabel={Odds Ratio (log scale)},
    ytick=data,
    yticklabels={
        $\Delta$LOC,
        Files,
        Tests,
        ForcePush,
        HasReview,
        HrsToReview,
        Copilot,
        Cursor,
        Devin,
        OpenAI
    },
    y=0.42cm,                     
    enlarge y limits=0.02,        
    legend style={
        at={(0.02,0.98)},
        anchor=north west,
        font=\scriptsize,
        fill=none,
        draw=none
    },
    label style={font=\scriptsize},
    tick label style={font=\scriptsize},
    xmajorgrids=true,
    grid style=dashed,
    scatter/classes={
        A={mark=*,mark size=2.2pt,blue},
        B={mark=square,mark size=2.4pt,draw=red!80!black,fill=white}
    }
]

\addplot[scatter,only marks,scatter src=explicit symbolic]
coordinates {
    (0.893,1) [B]
    (1.046,2) [B]
    (1.011,3) [B]
    (0.802,4) [B]
    (3.875,5) [B]
    (1.213,6) [B]
    (0.273,7) [B]
    (0.942,8) [B]
    (0.375,9) [B]
    (2.941,10) [B]
};

\addplot[red!80!black,thick] coordinates {(0.828,1) (0.963,1)};
\addplot[red!80!black,thick] coordinates {(0.938,2) (1.166,2)};
\addplot[red!80!black,thick] coordinates {(0.778,3) (1.314,3)};
\addplot[red!80!black,thick] coordinates {(0.662,4) (0.975,4)};
\addplot[red!80!black,thick] coordinates {(2.829,5) (5.308,5)};
\addplot[red!80!black,thick] coordinates {(1.127,6) (1.307,6)};
\addplot[red!80!black,thick] coordinates {(0.168,7) (0.442,7)};
\addplot[red!80!black,thick] coordinates {(0.547,8) (1.625,8)};
\addplot[red!80!black,thick] coordinates {(0.213,9) (0.661,9)};
\addplot[red!80!black,thick] coordinates {(1.628,10) (5.309,10)};

\addplot[scatter,only marks,scatter src=explicit symbolic]
coordinates {
    (0.890,1) [A]
    (1.039,2) [A]
    (1.006,3) [A]
    (0.797,4) [A]
    (3.830,5) [A]
    (1.205,6) [A]
    (0.278,7) [A]
    (0.956,8) [A]
    (0.375,9) [A]
    (2.893,10) [A]
};

\addplot[blue,thick] coordinates {(0.827,1) (0.959,1)};
\addplot[blue,thick] coordinates {(0.935,2) (1.153,2)};
\addplot[blue,thick] coordinates {(0.775,3) (1.305,3)};
\addplot[blue,thick] coordinates {(0.658,4) (0.966,4)};
\addplot[blue,thick] coordinates {(2.78,5) (5.25,5)};
\addplot[blue,thick] coordinates {(1.120,6) (1.298,6)};
\addplot[blue,thick] coordinates {(0.172,7) (0.449,7)};
\addplot[blue,thick] coordinates {(0.557,8) (1.639,8)};
\addplot[blue,thick] coordinates {(0.215,9) (0.658,9)};
\addplot[blue,thick] coordinates {(1.642,10) (5.094,10)};

\addplot[dashed] coordinates {(1,0) (1,11)};

\legend{Model A, Model B}

\end{axis}
\end{tikzpicture}
\vspace{-0.6cm}
\caption{Forest plot of logistic regression predictors}
\label{fig:forestplot}
\end{figure}
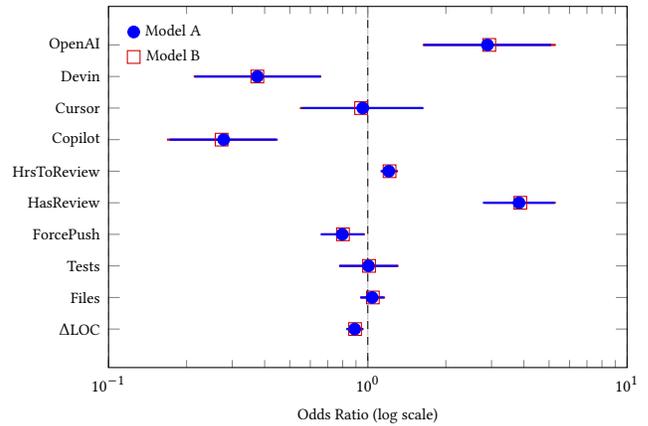

%% file: sections/figures/rq2_qualitative.tex
\begin{table}[t]
\centering
\footnotesize
\caption{Qualitative outcomes of agent-authored pull requests}
\label{tab:rq2_qualitative_summary}
\vspace{-0.3cm}
\begin{tabularx}{\linewidth}{>{\raggedright\arraybackslash}Xcccc}
\toprule
\textbf{Primary Driver} &
\textbf{\#} &
\textbf{Merged} &
\textbf{Not Merged} &
\textbf{Primary Outcome} \\
\midrule
Actionable review loop
& 32 & 30 & 2 & Success \\
Design disagreement
& 10 & 0 & 10 & Failure \\
Incomplete solution
& 7 & 0 & 7 & Failure \\
Process or policy issue
& 3 & 0 & 3 & Failure \\
Coordination break
& 2 & 0 & 2 & Failure \\
Incorrect or failing CI
& 2 & 0 & 2 & Failure \\
Other (scope, style, or stalled discussion)
& 4 & 0 & 4 & Failure \\
\midrule
\textbf{Total}
& \textbf{60} & \textbf{30} & \textbf{30} & \\
\bottomrule
\end{tabularx}
\end{table}

%% file: sections/discussion.tex
Across RQ1 and RQ2, our results show that integration of agent-authored pull requests is shaped primarily by review-time coordination rather than iteration volume. Quantitative models identify reviewer engagement and coordination stability as dominant signals, while the qualitative analysis explains how convergence through actionable feedback leads to successful merges. Conversely, design misalignment, incomplete solutions, and coordination disruptions consistently explain rejection.These findings highlight the need for alignment with established review practices.

%% file: sections/related.tex
\vspace{-4px}
\section{Related Work}
\label{sec:related}
Prior research characterizes pull request evaluation as a socio-technical
process in which integration depends on both technical and social factors.
Studies show that characteristics such as change size influence acceptance,
while review-time interaction plays a central role in integration decisions
\cite{gousios2014pullbased,rigby2014peer,tsay2014discussion}. Reviewer engagement,
discussion dynamics, and coordination practices help reviewers assess trust,
coordination cost, and perceived risk, often beyond code correctness alone.

Recent advances in large language models have enabled autonomous coding agents
to submit pull requests at scale. The AIDev dataset provides the first systematic
evidence of this shift, documenting adoption patterns and aggregate merge
outcomes \cite{aidev_zenodo_v3,li2025aidev}. However, existing studies largely
treat agent-authored pull requests as technical artifacts, leaving open how they
participate in established socio-technical review processes. Building on this
literature, our study examines whether review-time collaboration signals that
shape human pull request integration similarly influence agent-authored
contributions.

%% file: sections/threats.tex
\vspace{-4px}
\section{Threats to Validity}
\label{sec:limitations}

\emph{Construct validity.}
We infer collaboration from GitHub artifacts (e.g., reviews, commits, force pushes), which may not fully capture reviewer intent or interaction quality~\cite{kalliamvakou2016perils}. We mitigate this through complementary qualitative coding using a predefined codebook and discussion-based resolution of disagreements.

\emph{Internal validity.}
The study is observational and does not support causal claims. Although we control for agent identity and repository effects, unobserved project norms or reviewer practices may influence outcomes.

\emph{External validity.}
Our analysis focuses on agent-authored pull requests in large public GitHub repositories; findings may not generalize to private projects, smaller communities, or future agents with different behaviors.

%% file: sections/conclusion.tex
\vspace{-4px}
\section{Conclusion}
\label{sec:conclusion}

This paper studied how agent-authored pull requests integrate into human-driven development workflows using the AIDev dataset. We show that integration success depends primarily on review-time collaboration signals rather than iteration volume. Reviewer engagement increases merge likelihood when it leads to actionable feedback and convergence, while larger changes and coordination-disrupting behaviors reduce success. These findings highlight that effective agentic software engineering requires alignment with established code review and coordination practices.

To facilitate verification and replication of this work, a replication package~\cite{repkg_aidev_msr2026_anon} is made publicly available.